\documentclass[a4paper,12pt]{article}
\usepackage{epsfig}
\usepackage{times}
\usepackage{graphicx}
\usepackage{amsmath}
\usepackage{color}
\ifx\pdfoutput\undefined
\DeclareGraphicsExtensions{.epsi,.eps}
\else
\DeclareGraphicsExtensions{.pdf}
\fi
   
\newcommand{\mycite}[1]{\cite{#1}}

\newcommand{\beq}{\begin{equation}}
\newcommand{\eeq}{\end{equation}}
\newcommand{\beqa}{\begin{eqnarray}}
\newcommand{\eeqa}{\end{eqnarray}}

\def\opone{\leavevmode\hbox{\small1\kern-3.8pt\normalsize1}}

\begin{document}	
	\baselineskip24pt \noindent{\Large\bf Non-Hermitian Boundary State Distillation with Lossy Waveguides}\\[3mm]
	\noindent {\small \bf Walid Cherifi, Johan Carlstr\"om, Mohamed Bourennane$^*$ \& Emil J. Bergholtz}\\[2mm]
	\noindent {\it Department of Physics, Stockholm University, S-10691, Stockholm, Sweden}\\[2mm]

	%%%%%%%%%%%%%%%%%%%%%%%%%%%%%%%%%%%%%%%%%%%%%%%%%%%%%%%%%%%%%%%%%
	
	\noindent \textbf{The hallmark of topological phases is their exotic boundary states. In a series of remarkable experiments it has been shown that classical analogues of these states can be engineered in arrays of coupled optical waveguides given delicately fine-tuned input light.
	Here, we introduce and experimentally demonstrate a radically different approach in which a pattern of lossy waveguides distills the boundary states for generic, even defocused, input light, thus fully alleviating the need for fine-tuning. Our "topological distillation" approach is remarkably general: the lossy waveguides amount to an effectively non-Hermitian Hamiltonian, and the corresponding time-evolution (propagation of the light in the waveguides) removes the mundane bulk states of any topological (or trivial) band structure while retaining the intriguing boundary states by virtue of being the unique states with the longest life-time. We experimentally demonstrate the power and versatility of our approach by distilling the edge states in photonic graphene, as well as corner and edge states in non-Hermitian Kagome arrays. }  

%%%%%%%%%%%%%%%%%%%%%%%%%%%%%%%%%%%%%%%%%%%%%%%%%%%%%%%%%%%%%%%%%
%                           Introduction                        %
%%%%%%%%%%%%%%%%%%%%%%%%%%%%%%%%%%%%%%%%%%%%%%%%%%%%%%%%%%%%%%%%%
\newpage
\section*{Introduction}

Topologically protected boundary states are currently the focus of intense research, both in the context of fundamental research and applications. These states arise on surfaces, edges and corners of materials with a topologically nontrivial band structure \mycite{hasan2010colloquium, qi2011topological,weyl,Fu2010,benalcazar2017quantized,langbehn2017reflection,schindler2018higher,ezawa2018higher}. Due to being topologically protected, they are robust against perturbations as their existence does not hinge on microscopic details of the system. Beside their fundamental importance, these surface states have demonstrated considerable potential for applications in information technology \mycite{mellnik2014,Kondou}.

While much of the focus was initially directed at electronic systems, it has since been realized that photonic systems such as optical wave guide arrays allow for experimental realizations of topological systems \mycite{floquettopopt,lujoannopoulossoljacic,ozawa} that posses distinctive advantages over conventional materials: In optical systems it becomes possible to design lattice systems with minute control over model parameters like the hopping integral. Measuring the light at the output facet provides a comprehensive picture of the wave function at a level of detail that is impossible to achieve in electronic systems. 

The pertinent lattices of optical waveguides studied in this work are fabricated by carving in a glass medium using a fs laser \mycite{davis1996writing}. In this process, the refractive index is altered along lines the run through the sample, forming waveguides along which the light tends to propagate. Placing waveguides in close proximity, the light has a finite amplitude to swap between them in a process that is analogous to tunneling in an electronic system. The propagation of light through a waveguide array is described by the paraxial equation, whose structure is identical to the Schr\"odinger equation: Identifying the axis along which the light travels with time, the optical setup directly emulates the dynamics of a quantum system despite being classical. 
Key phenomena that have been observed in optical wave guides include chiral edge states \mycite{floquettopopt,maczewsky2017observation}, corner states \mycite{noh2018observation,hasan2019}, defect modes in distorted honeycomb arrays \mycite{noh2018observation}, localized bulk modes in Lieb lattice arrays \mycite{mukherjee2015observation,lieb2} as well as unconventional phase transitions \cite{zeuner2015,szameit2022} and band structures \mycite{cerjan2019,leykam2017}. 

Optical waveguide arrays also permit a unique degree of freedom that is not present in electronic systems: By introducing cuts in the waveguides, the light is compelled to partially scatter, and leak out of the system. In the context of the quantum mechanics analogy, this translates to a Schr\"odinger equation which features a non-Hermitian Hamiltonian and correspondingly a non-unitary time-evolution \cite{elganainy,El-Ganainy2019,ozdemir}. Non-Hermitian topology is an extremely rich field \mycite{gong,NHreview} with phenomena ranging from exceptional points \cite{elganainy,ozdemir,miri} to open Fermi surfaces \mycite{NHarc,carlstrom2018} and the skin effect \mycite{skinreview1,skinreview2}. Optical and photonic systems is now the main platform for observing these effects \mycite{El-Ganainy2019,miri}, with examples harnessing wave guide arrays specifically ranging from topological phase transitions \mycite{zeuner2015,szameit2022} to exceptional rings in the spectrum \mycite{cerjan2019}.

In this work we exploit non-Hermitian effects to distill topological boundary states: By applying dissipation to one or two sub-lattices of the system, any eigenstate that has a presence on those must necessarily posses a finite life time. Crucially, the only states which are confined to a single sub-lattice are the boundary states, and our setup provides a prescription for removing all other components of the wave-function. The protocol of topological distillation allows boundary states to be isolated to exponential precision, irrespectively of the input light.

\section*{Emulating quantum lattice models with light}
In the pertinent optical waveguide array setup, Maxwells equations, which describe propagating light are captured by the paraxial equation:
\begin{equation}i\frac{d}{dz}\phi(x,y,z)=\left ( -\frac{1}{2k_0}\nabla^2_{x,y}-\frac{k_0}{n_0}\delta n(x,y)\right )\phi(x,y,z), \label{parax}
\end{equation} 
where, $\phi$ describes the electric field envelope function, which is related to the electric field according to $\mathbf{E}(x,y,z)=\phi(x,y,z)\exp(i k_0 z-i\omega t)\mathbf{\hat n}$. Here, $n_0$ denotes the refractive index of the host materials, while $k_0=2\pi n_0/\lambda$ is the corresponding wavenumber.
The function $\delta n(x,y)$ describes a deviation of the refractive index from that of the uniform host material, $n_0$. 

Identifying the $z-$direction  of eq. (\ref{parax}) with the time-axis, it becomes identical to the Schr\"odinger equation with $\hbar=1$. 
An effective potential is obtained from a spatially inhomogeneous refractive index as described by the function $\delta n(x,y)$. Using a fs laser to carve the waveguides, the effective potential attains sharp minima, allowing the systems to be accurately described as a tight-binding model. The effective parameters of this model depend on details of the manufacturing, as well as the wavelength of the injected light.

By introducing  cuts in a waveguide, the propagating light is compelled to partially scatter, effectively rendering the system dissipative. At the level of a tight-binding description, this results in an imaginary chemical potential, implying a non-Hermitian Hamiltonian on the form: 
\begin{equation}
H=\sum_{ij}t_{ij} c_i^\dagger c_j- \sum_i i\gamma_i c_i^\dagger c_i,\;\;\; \gamma_i\ge 0.
\end{equation} 
In this work, we use this method to engineer non-Hermitian tight-binding models with staggered dissipation, focusing on the kagome and anisotropic honeycomb lattices as illustrated in Fig. \ref{fig:overview}. The kagome lattice exhibits two lattice spacings, $d_1$ (dashed lines) and $d_2$ (solid lines), resulting in two effective tunneling amplitudes $t_\alpha\sim e^{-d_\alpha/\xi}$, where $\xi$ is a characteristic length scale that typically depends on the optical wavelength. The honeycomb lattice is normally described by a single lattice spacing $d$ and a tunneling amplitude $t$. However, we consider an anisotropic honeycomb lattice to compensate for the fact that our waveguides are not rotationally symmetric, see discussion below. 
We assume that these parameters are sufficient to capture the essential physics of the system, and this is also indicated by our experiments as discussed below.

The kagome lattice in Fig. \ref{fig:overview}, a-b, supports zero-energy corner or edge states, depending on the tunneling amplitudes. For $t_1<t_2$, a single corner state \mycite{ezawa2018higher,kunst2018lattice} is bound to the top of the triangle. When $t_1>t_2$, a zero-energy mode is localized to the bottom edge. The honeycomb lattice in Fig. \ref{fig:overview}, c, supports a family of zero-energy modes that are localized to the upper edge.
 
 By the introduction of dissipative waveguides, the time-evolution of the system becomes effectively non-unitary as light can leak from the system. Here, we consider the case when one or two sub-lattices are dissipative, so that any state which has a finite amplitude on these is exponentially suppressed for larger values of $z$, which is the equivalent of time in the Schr\"odinger equation. However, the lattice geometries we consider are chosen such that the only states which are confined to the non-dissipative sub-lattices are the corner and edge states, the remaining modes all have an overlap with the dissipative waveguides. In fact, such lattice geometries can be found for essentially any topological (or non-topological) boundary state as can be inferred from earlier studies of Hermitian tight-binding models with open boundaries \cite{kunst2018lattice,kunst2018boundaries}. Consequently, as light propagates through the waveguide array, it is subject to a ``distillation'' process where all modes except for the edge states are removed. In the end, we are left with only the "topologically" nontrivial part. We stress that this procedure is in principle exact in long time limit as long as the short-range tight-binding description holds. Here it should be noted that for symmetry protected states, the loss as well as the lattice geometry giving the exact steady states is explicitly symmetry breaking. Nevertheless the distilled steady states genuinely represent the symmetry protected boundary states. In the case of the kagome corner states this situation is particularly intricate and may be cast in terms of sub-symmetry protected topological states \cite{Buljan}. If instead applied to symmetric lattices, states at a given boundary are targeted while oppositely localised states in contrast exhibit the shortest life time of all.

%\section{Methods:}
%\subsection{Fabrication of the photonic structures:}

\section*{Experiment}
To realize the aforementioned lattice model
we use photonic waveguide arrays that are fabricated using a pulsed fs laser (BlueCut fs laser from Menlo Systems). The fs laser produces light pulses centered at a wavelength of $ 1030$ $ $$nm$, having a  duration of $ 350 $ $ fs$. In this experiment, we use a repetition of $ 1 $  $MHz$. The waveguides are written in a Corning EAGLE2000 alumino-borosilicate glass sample with dimensions ($L= 50, W = 25, h =1.1$) $ mm $. To inscribe the waveguide structures, pulses of $ 210$ $ nJ$  are focused using a $ 50 $ $X$ objective with a numerical aperture (NA) of $ 0.55$. The waveguides are written at a depths between $70$ to $175$ $ \mu $$ m $ under the surface according to the designed structure, while the sample is translated at a constant speed of 30 mm/s by a high-precision three axes translation stage (A3200, Aerotech Inc.). The fabricated waveguides support a Gaussian single mode at $780$ $nm$, with a mode field diameter ($1/e^2$) of approximately $ 6-8 $ $\mu $$m$. %The mechanism of ultrafast laser pulses-material interaction give refractive index increase of about $ 2\times 10^{-4}$. 
The propagation losses are estimated to be around $ 0.3$ $dB/cm $ and the birefringence in the order of $ 7 \times 10^{-5}$.
After the structures are written in the glass sample, the lateral facets are carefully polished down to an optical quality with a roughness of $ 0.1$ $\mu$$m$. The sample is examined using a microscope, where each individual waveguide input, output, and position is verified in accordance with the designed structure. 

At the input facet, we inject a coherent light beam from a tunable laser (Cameleon Ultra II, Coherent) using a $100\times$ objective of $0.9$ NA, which is sufficient for individual excitation of all waveguides composing the  structure. The output light is collected with the help of a $100\times $, high NA objective and imaged using a CCD camera.

%These arrays were written by a focused fs laser beam on $50$ $mm$ length glass samples that were mounted on 3 axis nano-positing stages. The waveguides were written such as to support a single mode at $780$ $nm$. {\color{red} Dissipation was introduced by creating cuts... }

\begin{figure}[ht]
	\centering
	\begin{tabular}{ccc}
	\includegraphics[width=1.0\linewidth]{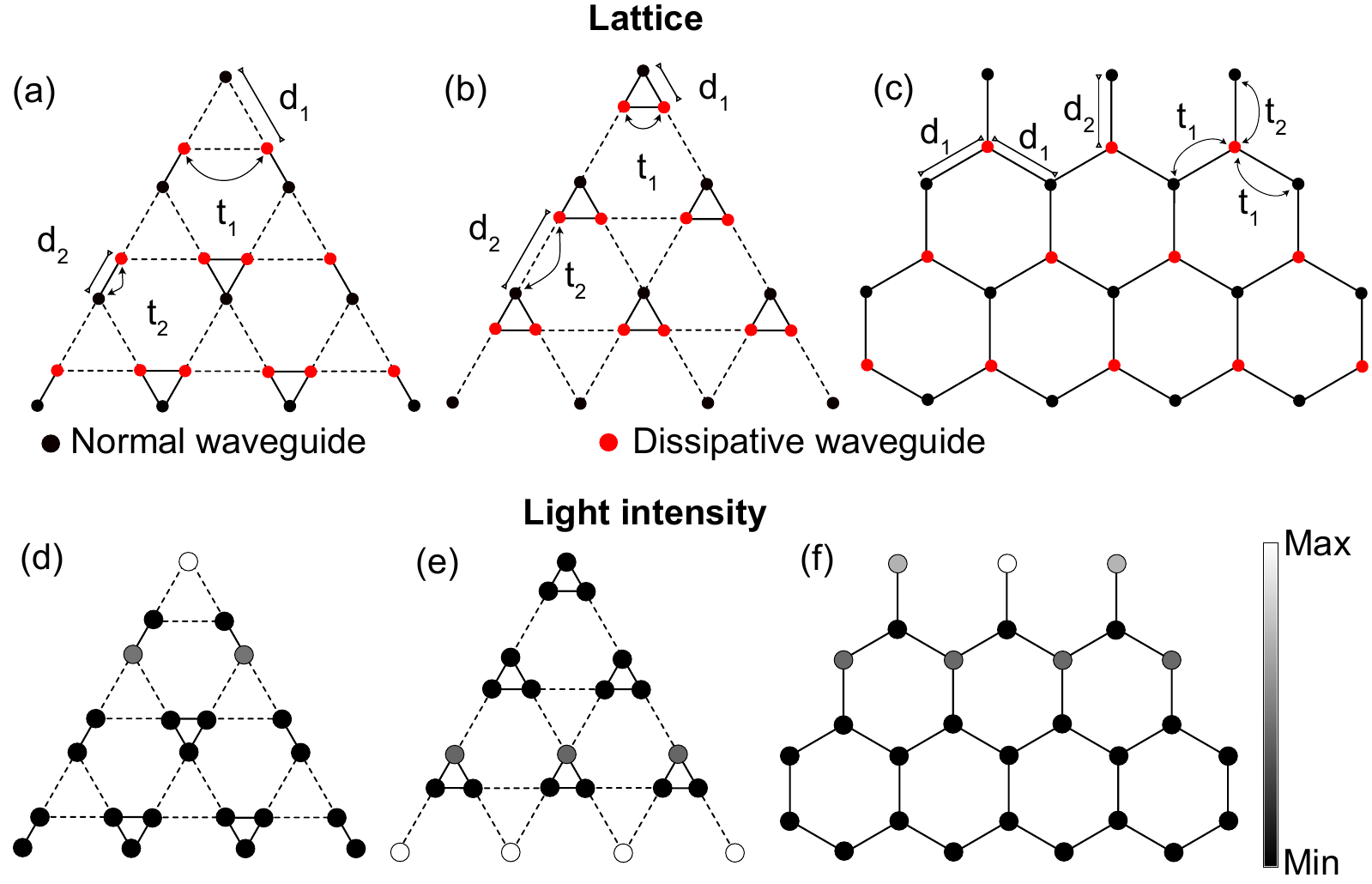}
	\end{tabular}
	\caption{{\bf Lattice structures and effective models.} The Kagome lattice (a-b) is characterized by two lattice constants, $d_1$ and $d_2$, with corresponding tunneling amplitudes, $t_1$ and $t_2$. For $d_1>d_2$ (a), we find that $t_1<t_2$, since tunneling is more prevalent when the waveguides are in closer proximity. When $d_1<d_2$ (b), we instead observe that $t_1>t_2$. 
	The honeycomb lattice (c) is made to be anisotropic, with two lattice spacings	to compensate for the shape of the waveguides: Because these are not rotationally symmetric, we obtain $t_2>t_1$ for an isotropic lattice. By taking $d_2>d_1$, this effect is mitigated.  
	A fraction of the waveguides are made to be dissipative, thus giving rise to effectively non-Hermitian terms in the paraxial equation (\ref{parax}). As a result, any state which overlaps with these is subject to exponential suppression. Asymptotically, this leads to a 
	``distilled'' state where only the topological corner or edge states survive, as illustrated in (d-f). The Kagome lattice (a) gives rise to the corner state (d), while (b) results in an edge state (e). The honeycomb lattice (c) hosts several zero energy states confined to the edge, which together produce the amplitude (f). 
	} 
	\label{fig:overview}
\end{figure}

\begin{figure}[ht]
	\centering
	\begin{tabular}{ccc}
	\includegraphics[width=0.5\linewidth]{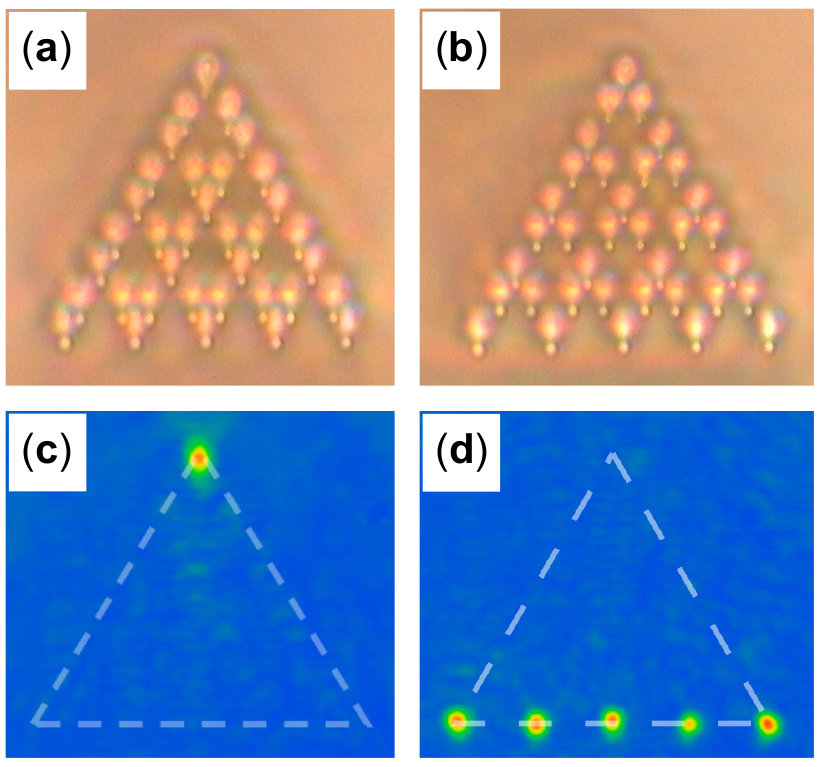}
	\end{tabular}
	\caption{{\bf  Observation of corner and edge states in two Kagome lattices.} 
	(a) Microscope image of a Kagome lattice with $d_1=12 \mu m$ and $d_2=9 \mu m$. Since $d_1> d_2$, this corresponds to the scenario of Fig. \ref{fig:overview} a, and we expect a corner state that is localized to the top of the triangle. 
	(b) Microscope image of a Kagome lattice with $d_1=9 \mu m$ and $d_2=\mu m$. Here, $d_1 < d_2$, as illustrated in Fig. \ref{fig:overview} b, implying an edge state at the bottom of the triangle. 
 (c) Intensity of the light emerging from the waveguide array (a), as captured by the CCD camera. The injected light is defocused, and thus enters all of the waveguides in the array. 
	The dissipative waveguides remove all modes with which they overlap, leaving only the corner state at the output facet. 
	(d) Intensity of the light emerging from the array (b). The dissipative waveguides remove all modes except the edge state. 
		 The wavelength of the injected light is $780 nm$. } 
	\label{fig:kago}
\end{figure}

To realize a kagome lattice as shown in Fig. \ref{fig:overview} a, we use a sample with $d_1=12\mu m$ and $d_2=9\mu m$ such that in the limit $\xi\ll d_i$, the system can be described with two tunneling parameters $t_1<t_2$. Two of the three sub-lattices are made dissipative as illustrated. In this setup we expect to isolate a state confined to the upper corner of the triangle as shown in Fig. \ref{fig:overview}, d. The measured light intensity, displayed in Fig. \ref{fig:kago} (a,c), confirms this expectation. At the output facet we see a single strong light beam propagating through the waveguide at the top of the triangle. Subsequent waveguides do not carry enough light for it to be registered. Noting that the corner state is exponentially localized as $|\psi_n|\sim (t_1/t_2)^n$ (with $n$ denoting the distance from the corner), the strong localization of the corner state is consistent with $t_1 \ll t_2$. This disparity can in turn be understood from the fact that the tunneling amplitude decays exponentially as $t_\alpha\sim e^{-d_\alpha/\xi}$.

The second type of kagome lattice, shown in Fig. \ref{fig:overview} b, requires $d_1<d_2$ so that $t_1>t_2$. In this scenario, we expect to find a state that is exponentially localized to the lower edge as seen in Fig. \ref{fig:overview} e. To realize this class of system, we use a sample with $d_1=9\mu m$ and $d_2=12\mu m$, in which we inject light with a wavelength of $ 780 nm$. As in the previous case, two of the sub-lattices are made dissipative. The outcome of the experiment is summarized in Fig. \ref{fig:kago}, (b,d). Despite that light is injected in all waveguides, we observe that it is confined to the lower edge of the sample at the output facet. Except for some stray light, we do not register a finite amplitude elsewhere in the sample. Noting that the edge state, like the corner state, is exponentially localized, this is consistent with a scenario where there is a large disparity in tunneling amplitude so that $t_1\gg t_2$.

\begin{figure}[ht]
	\centering
	\begin{tabular}{ccc}
	\includegraphics[width=0.5\linewidth]{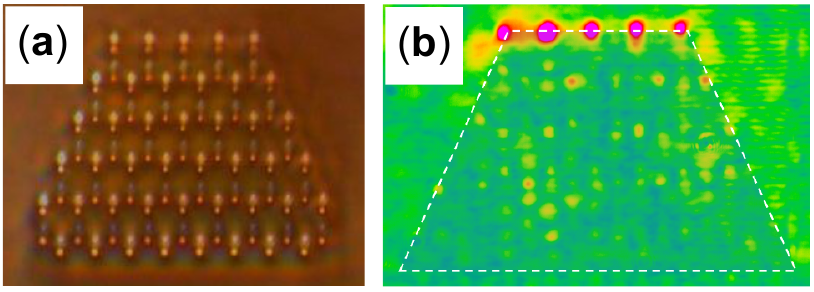}
	\end{tabular}
	\caption{{\bf  Observation of edge state in an anisotropic honeycomb lattice} of waveguides with lattice spacings $d_1=9\mu m$ and $d_2=13\mu m$ respectivly.
	 (a), Microscope image of the honeycomb lattice. (b), Intensity of the light emerging from the waveguides, as captured by the CCD camera. The injected light is defocused, and thus enters all of the waveguides in the array. The honeycomb lattice exhibits a family of gap-less edge states that are confined to one sub-lattice. All other modes are removed by the dissipative waveguides, leaving only the edge illuminated--see also Fig. \ref{fig:overview}, d and g. 
	 The wavelength of the injected light is $780n m$. 
	} 
	\label{fig:honey}
\end{figure}

The anisotropic honeycomb lattice shown in Fig. \ref{fig:overview} c, is characterized by two lattice spacings $d_1$, $d_2$. Assuming that the tunneling length scale $\xi$ is small by comparison, this system can be modeled by two tunneling amplitudes $t_1$, $t_2$. If $d_1=d_2$, then we expect that $t_1>t_2$, due to the anisotropic nature of the waveguides. To compensate for this fact, we use an 
 array where $d_1>d_2$, thus reducing $t_1$. 
To realize this scenario, we use a sample with $d_1 = 9\mu m,\; d_2 = 13\mu m $, and inject light with a wavelength of $\lambda=780 nm$. One of the two sub-lattices is made dissipative. This lattice type supports a family of edge state that are confined to the upper edge of the system, which do not overlap with the dissipative sub-lattice. Injecting defocused light that activates all waveguides, we expect that all modes except for the edge states are removed as the light propagates through the sample, as illustrated in Fig. \ref{fig:overview} f. The outcome of the experiment is shown in Fig. \ref{fig:honey}. At the output facet, we clearly see that the light is confined to the upper edge.

\section*{Discussion}
In this work, we have provided three examples of how dissipative waveguides can be used to isolate nontrivial edge states through a process of ``distillation''. In the language of the Schr\"odinger equation, the system is described by a non-Hermitian Hamiltonian, which facilitates a non-unitary time-evolution. The defocused light which is injected into the system lacks any structure and is agnostic to the specific edge states supported by the lattice. As it evolves, order develops in the system. 

While our experiment relies on classical light, the underlying concept is still applicable in the quantum realm, despite the conceptual challenges facing this field: 
Although open quantum systems typically undergo an effectively non-unitary time-evolution, they cannot, in the general case, be identified with a non-Hermitian Hamiltonian, despite glaring phenomenological similarities. This fact notably limits the applicability of non-Hermitian models to actual quantum systems.
Nonetheless, the concept of ``distilling'' edge and corner states directly carries over to quantum systems. As illustrated in our experiment, the remaining states can be vacated by introducing a ``dissipative'' sub-lattice where particles can escape the system. This is a rare example of when non-Hermitian models have a precise predicting power in an open quantum system.

Our distillation approach is remarkably general can be harnessed for robustly creating essentially any boundary state by choosing appropriate lattices and loss patterns. The general recipe for this is both powerful and straigthforward: we consider lattices that harbour boundary states are locally unique eigenstates with zero amplitude on certain sublattices \cite{kunst2018lattice,kunst2018boundaries}---and let the corresponding sites be represented by lossy waveguides. This procedure is in principle asymptotically exact at long times (long waveguides) independently of the input light as long as it has some finite overlap with the target state(s). This hence alleviates the key issue of fine-tuning the input light and thus opens up new avenues for both fundamental studies and possible applications of photonic topological boundary states.

%% Put the bibliography here, most people will use BiBTeX in
%% which case the environment below should be replaced with
%% the \bibliography{} command.

%% Put the bibliography here, most people will use BiBTeX in
%% which case the environment below should be replaced with
%% the \bibliography{} command.

\begin{thebibliography}{10}

\bibitem{hasan2010colloquium}
Hasan, M. Z. \& Kane, C. L. Colloquium: Topological insulators. \textit{Rev. Mod. Phys.} {\bf 82}, 3045 (2010).

\bibitem{qi2011topological}
Qi, X.-L. \& Zhang, S.-C. Topological insulators and superconductors. \textit{Rev. Mod. Phys.} {\bf 83}, 1057 (2011).

%\bibitem{hosur2013}
%Hosur, P. \& Qi, X. Recent developments in transport phenomena in Weyl semimetals. \textit{C. R. Phys.}  {\bf 14} 857 (2013).

\bibitem{weyl} Armitage, N.P., Mele, E.J.  \& Vishwanath, A., Weyl and Dirac semimetals in three-dimensional solids. %\textit{Rev. Mod. Phys.} {\bf 90}, 015001 (2018).

\bibitem{Fu2010}
Fu, L. Topological Crystalline Insulators. \textit{Phys. Rev. Lett.} {\bf 106}, 1068-2 (2011).

\bibitem{benalcazar2017quantized}
 Benalcazar, W. A., Bernevig, B. A. \& Hughes, T. L. Quantized electric multipole insulators. \textit{Science} {\bf 357}, 61 (2017).

\bibitem{langbehn2017reflection}
Langbehn, J., Peng, Y., Trifunovic, L., von Oppen, F. \& Brouwer. P. W. Reflection-Symmetric Second-Order Topological Insulators and Superconductors. \textit{Phys. Rev. Lett.} {\bf 119}, 246401 (2017).

\bibitem{schindler2018higher}
Schindler, F., Cook, A. M., Vergniory, M. G., Wang, Z., Parkin, S. S. P., Bernevig,  B. A. \& Neupert, T. Higher-order topological insulators. \textit{Science Adv.} {\bf 4}, eaat0346 (2018).

\bibitem{ezawa2018higher}
Ezawa, M. Higher-order topological insulators and semimetals on the breathing Kagome and pyrochlore lattices. \textit{Phys. Rev. Lett.} {\bf 120}, 026801 (2018).
%----------



\bibitem{mellnik2014}
Mellnik, A. R. et al. Spin-transfer torque generated by a topological insulator. \textit{Nature} {\bf 511}, 449 (2014). 

\bibitem{Kondou}
Kondou, .K et al. Fermi-level-dependent charge-to-spin current conversion by Dirac surface states of topological insulators. \textit{Nat. Phys. } {\bf 12}, 1027 (2016).



%--------------

\bibitem{floquettopopt}
Rechtsman, M. C., Zeuner, J. M., Plotnik, Y., Lumer, Y., Podolsky, D., Dreisow, F., Nolte, S., Segev, M.  \& Szameit, A. Photonic Floquet topological insulators. \textit{Nature} {\bf 496}, 196 (2013).

\bibitem{lujoannopoulossoljacic}
Lu, L., Joannopoulos, J.D., \& Solja\v{c}i\'{c}, M. Topological photonics. \textit{Nature Photonics} {\bf 8}, 821 (2014).

\bibitem{ozawa} Ozawa, T. et al. Topological photonics. \textit{Rev. Mod. Phys.} {\bf 91}, 015006 (2019).

\bibitem{davis1996writing}
Davis, K. M., Miura, K., Sugimoto, N. \& Hirao, K. Writing waveguides in glass with a femtosecond laser. \textit{Opt. Lett.} {\bf 21}, 1729 (1996).



 
\bibitem{maczewsky2017observation}
Maczewsky, L. J., Zeuner, J. M., Nolte, S. \& Szameit, A. Observation of photonic anomalous Floquet topological insulators. \textit{Nature Commun.} {\bf 8}, 13756 (2017).
 
 \bibitem{noh2018observation} Noh, J.,  Benalcazar, A. W., Huang, S., Collins, M. J., Chen, K. P., Hughes, T. L. \& Rechtsman, M. C.  
Topological protection of photonic mid-gap defect modes.
\textit{Nature Photonics} {\bf 12}, 408 (2018).

\bibitem{hasan2019} El Hassan, A., Kunst, F.K., Moritz, A., Andler, G., Bergholtz, E.J. \& Bourennane, M. 
Corner states of light in photonic waveguides.
\textit{Nature Photonics}  {\bf 13}, 697 (2019).
 
\bibitem{mukherjee2015observation}
Mukherjee, S., Spracklen, A., Choudhury, D., Goldman, N., \"{O}hberg, P., Andersson, E. \& Thomson, R. R. Observation of a Localized Flat-Band State in a Photonic Lieb Lattice. \textit{Phys. Rev. Lett.} {\bf 114}, 245504 (2015).

\bibitem{lieb2}
Vicencio, R.A., Cantillano, C., Morales-Inostroza, L., Real, B., Mejia-Cortes, C., Weimann, S., Szameit, A. and Molina, M. I. Observation of Localized States in Lieb Photonic Lattices. \textit{Phys. Rev. Lett.} {\bf 114}, 245503 (2015).



\bibitem{zeuner2015}
Zeuner, J. M. et al. Observation of a Topological Transition in the Bulk of a Non-Hermitian System. \textit{Phys. Rev. Lett.} {\bf 115}, 040402 (2015).

\bibitem{szameit2022}
Widemann, S. \& Kremer, M. \& Longhi, S. \&Szameit, A. Topological triple phase transition in non-Hermitian Floquet quasicrystals. \textit{Nature} {\bf 601}, 354 (2022).

\bibitem{cerjan2019}
Cerjan, A., Huang, S., Wang, M., Chen, K.P., Chong, Y. \& Rechtsman, M.C. Experimental realization of a Weyl exceptional ring. \textit{Nature Photonics} {\bf 13} 623 (2019).

\bibitem{leykam2017}
Noh, J.,  Huang, S., Leykam, D., Chong, Chen Y.D. , K.P. \& Rechtsman, M.C. Experimental observation of optical Weyl points and Fermi arc-like surface states. \textit{Nature Photonics} {\bf 13} 611 (2017).



\bibitem{elganainy} El-Ganainy, R., Makris, K.G., Khajavikhan, M., Musslimani, Z.H., Rotter, S. \&  Christodoulides,  D.N.   Non-Hermitian physics and PT symmetry. \textit{Nature Physics} {\bf 14}, 11 (2018).

\bibitem{El-Ganainy2019}
El-Ganainy, R. \& Khajavikhan, M. \& Christodoulides, D. N. \& Ozdemir, S. K. The dawn of non-Hermitian optics.  \textit{Communications Physics} {\bf 2}, 37 (2019).

 \bibitem{ozdemir}  \"Ozdemir, S.K., Rotter, S., Nori, F., \& L. Yang, Parity-time symmetry and exceptional points in photonics.
\textit{Nature Materials} {\bf 18}, 783 (2019).


\bibitem{gong}
Gong, Z., Ashida, Y., Kawabata, K., Takasan, K., Higashikawa, S., \& Ueda, M.  Topological Phases of Non-Hermitian Systems, \textit{Phys. Rev. X} {\bf 8}, 031079 (2018).

\bibitem{NHreview}
Bergholtz, E.J., Budich, J.C. \&  Kunst,  F.K.  Exceptional topology of non-Hermitian systems, \textit{Rev. Mod. Phys.} {\bf 93}, 015005 (2021).


 \bibitem{miri} Miri, M.-A., \& Alu, A., Exceptional points in optics and photonics. \textit{Science}, {\bf 363} eaar7709 (2019).

%
%\bibitem{hassan2019}
%El Hassan, A. \& Kunst, F. K.  \& Moritz, A. \& Andler, G.  \& Bergholtz, E. \& Bourennane, M. \textit{Nature Photonics} {\bf 13} 697 (2019).

%NH PHYSICS%
%\bibitem{NHbook}
%Y. Ashida, Z. Gong, and M. Ueda, {\em Non-Hermitian Physics}, \href{https://dx.doi.org/10.1080/00018732.2021.1876991}{Advances in Physics 69, {\bf 249} (2020)}.


\bibitem{NHarc} Zhou, H., et. al., , Observation of bulk Fermi arc and polarization half charge from paired exceptional points, {\it Science} {\bf 359}, 1009 (2018).

\bibitem{carlstrom2018}
Carlstr\"om, J. \& Bergholtz, E. Exceptional links and twisted Fermi ribbons in non-Hermitian systems,  \textit{Phys. Rev. A} {\bf 98} 042114 (2018).


\bibitem{skinreview1} Okuma, N. \&  Sato, M. Non-Hermitian Topological Phenomena: A Review
\textit{Ann. Rev. of Cond. Mat. Phys.} {\bf 14} 83 (2023).

\bibitem{skinreview2}
Lin, R., Tai, T., Li, L., \&  Lee, C. H. Topological Non-Hermitian skin effect. arXiv:2302.03057.


\bibitem{kunst2018lattice}
Kunst, F. K., van Miert, G. \& Bergholtz, E. J. Lattice models with exactly solvable topological hinge and corner states. \textit{Phys. Rev. B} {\bf 97}, 241405(R) (2018).

\bibitem{kunst2018boundaries}
Kunst, F. K., van Miert, G. \& Bergholtz, E. J. Boundaries of boundaries: a systematic approach to lattice models with solvable boundary states of arbitrary codimension. \textit{Phys. Rev. B} {\bf 99}, 085426 (2019); Extended Bloch theorem for topological lattice models with open boundaries. \textit{Phys. Rev. B}   {\bf 99}, 085427 (2019).


\bibitem{Buljan} Wang, Z., Wang, X., Hu, Z., Bongiovanni, D., Jukic, D., Tang, L., Song, D., Morandotti, R.,  Chen, Z. \& Buljan, H. Sub-symmetry protected topological states,  arXiv:2205.07285.

%\bibitem{su1979solitons}
%Su, W. P., Schrieffer, J. R. \& Heeger, A. J. Solitons in Polyacetylene. \textit{Phys. Rev. Lett.} {\bf 42}, 1698 (1979).






\end{thebibliography}

\textbf{Acknowledgements}
This work is supported by the Swedish research council (grant 2018-00313), the Knut and Alice Wallenberg Foundation (grant 2018.0460), and as well as the G\"oran Gustafsson Foundation for Research in Natural Sciences and Medicine. We thank Naemi Florin for early discussions on the topic.\\
 
\textbf{Author Contributions}

E.J.B. initiated the research and conceived the general idea. V.C. designed and carried out the experiment. M.B. supervised the experimental part. J.C. and E.J.B. derived the theoretical results and wrote the main text. M.B. and V.C. wrote the experimental part. All the authors discussed the results and contributed to the final version of the manuscript.\\
  
\textbf{Competing interests statement} 

The authors declare that they have no competing financial interests.\\

\textbf{Correspondence}

Correspondence and requests for materials
should be addressed to M.B. ~(email: boure@fysik.su.se).

\newpage

\newpage

\begin{figure}[ht]
	\centering
	\begin{tabular}{ccc}
	\includegraphics[width=0.5\linewidth]{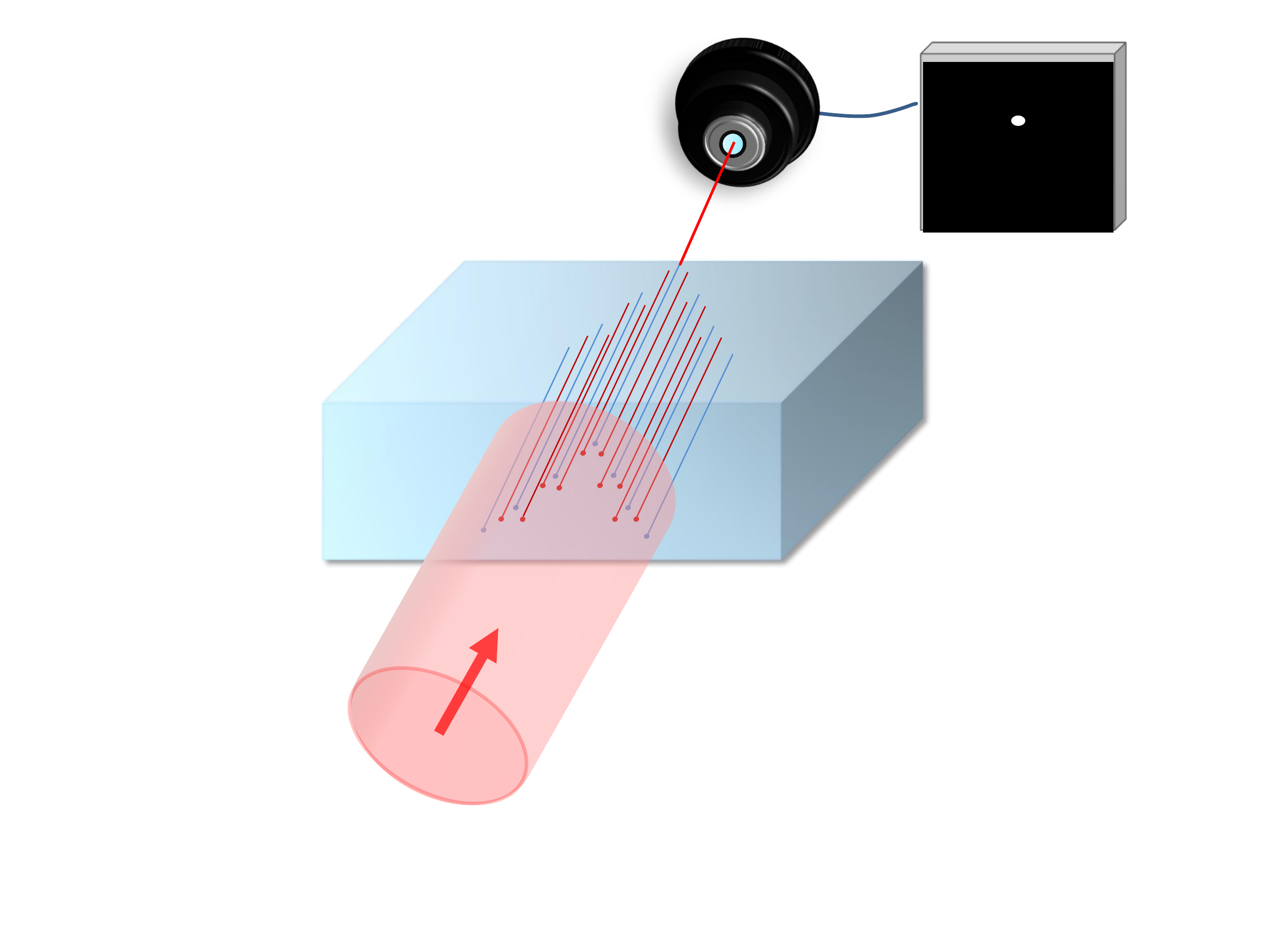}
	\end{tabular}
	\caption{{\bf Sketch of the experimental setup.} The coherent light beam from a tunable laser is launched into the glass sample using a $\times 100$ objective of $0.9$ NA,  which is sufficient for individual excitation of each waveguide (blue and red lines represent high quality and lossy waveguides respectively). The output light is collected with similar objective. A CCD camera was used to collect the image profile of each individual waveguide of the topological structure.
	} 
	\label{fig:experiment}
\end{figure}
\newpage

\newpage

\newpage

\end{document}